\documentclass[aoas,MSNbibl,nameyear,dvips]{arximspdf}

\doi{10.1214/11-AOAS485}
\volume{5}
\issue{2B}
\pubyear{2011}
\firstpage{1127}
\lastpage{1131}

\begin{document}
\begin{frontmatter}

\title{Special section on statistics in neuroscience}
\runtitle{Introduction}

\begin{aug}
\author[A]{\fnms{Karen} \snm{Kafadar}\corref{}\ead[label=e1]{kkafadar@indiana.edu}}
\runauthor{K. Kafadar}
\affiliation{Indiana University}
\address[A]{Department of Statistics \\
Indiana University \\
Bloomington, Indiana 47408-3825\\
USA\\
\printead{e1}} %adresu isvedimo komanda gale!
\end{aug}

% HISTORY:
\received{\smonth{5} \syear{2011}}

% ABSTRACT
\begin{abstract}
This article provides a brief introduction to seven papers
that are included in this special section on Statistics in
Neuroscience:\\[-5pt]

\noindent\hangindent=15.3pt
(1)\hspace*{5pt}Xiaoyan Shi, Joseph G. Ibrahim, Jeffrey Lieberman,
Martin Styner, Yimei Li and Hongtu Zhu: Two-state empirical
likelihood for longitudinal neuroimaging data

\noindent\hangindent=15.3pt
(2)\hspace*{5pt}Vincent Q. Vu, Pradeep Ravikumar, Thomas Naselaris,
Kendrick N. Kay, Jack L. Gallant and Bin Yu:
Encoding and decoding V1 fMRI responses to natural
images with sparse nonparametric models

\noindent\hangindent=15.3pt
(3)\hspace*{5pt}Sourabh Bhattacharya and Ranjan Maitra:
A nonstationary nonparametric Bayesian approach to
dynamically modeling effective connectivity in
functional magnetic resonance imaging experiments

\noindent\hangindent=15.3pt
(4)\hspace*{5pt}Christopher J. Long, Patrick L. Purdon, Simona
Temereanca, Neil U. Desai, Matti S. H\"{a}m\"{a}l\"{a}inen and
Emery Neal Brown: State-space solutions to the dynamic
magnetoencephalography inverse problem using high performance
computing

\noindent\hangindent=15.3pt
(5)\hspace*{5pt}Yuriy Mishchencko, Joshua T. Vogelstein and Liam Paninski:
A Bayesian approach for inferring neuronal connectivity from
calcium fluorescent imaging data

\noindent\hangindent=15.3pt
(6)\hspace*{5pt}Robert E. Kass, Ryan C. Kelly and Wei-Liem Loh:
Assessment of synchrony in multiple neural spike trains
using loglinear point process models

\noindent\hangindent=15.3pt
(7)\hspace*{5pt}Sofia Olhede and Brandon Whitcher: Nonparametric tests of
structure for high angular resolution diffusion imaging in Q-space

\end{abstract}

% KEYWORDS

\end{frontmatter}

\section{Introduction}

In a lecture at Indiana University in March 2008,
Peter Hall offered several valuable insights about
the field of statistics, three of which are noted below:
\begin{enumerate}
\item Advances in statistics have
come from the need to analyze different data types
(``Statistics is `reactive;' it is very responsive to
new problems that arise in chemistry, biology, physics, \ldots'').
\item Data sets continue to increase in size.
\item Computational algorithms are essential components of
the analysis: ``Advances in powerful computing equipment
has had a dramatic impact on statistical methods and theory.
It has changed forever the way data are analyzed.''
\end{enumerate}
The seven articles in this special section on
\textit{Statistics and Neuroscience}, together with two earlier \textit{AOAS} articles,
vividly illustrate all three principles.

Function of the human nervous system has fascinated researchers
for decades, due to its complex network of interactions among
critical parts of its components in the central nervous system
(brain, spinal cord, retina) and periphery (nerves).  The
amount of data that can be collected on these individual
components is truly massive, now that instruments for
measuring signals (responses to stimuli) have been developed
with increasing resolution (spatially and temporally) and
sensitivity (weaker signals in the presence of high noise
levels).  The range of statistical methods that are needed
to understand neural and brain development, functionality,
and interactions is extremely broad.  This special section
includes seven articles that present useful statistical
methodology designed to address various aspects of data that
arise in neuroscience, specifically with brain imaging data
collected via functional magnetic resonance imaging (fMRI)
or other imaging techniques, and the analysis of neural
spike train data.
The articles demonstrate the wide variety of statistical
problems, the diversity of methods that can be applied,
and, most importantly, the valuable insights that are
obtained through the application of sound statistical
methods.

Functional magnetic resonance imaging was developed in the
early 1990s for brain imaging [e.g., Ogawa et al. (\citeyear{Ogaetal92})] and
immediately presented statisticians with a huge new area of
problems to be considered: the analysis of massive data sets.
The data, changes in blood flow in response to neural activity
[blood oxygen level dependent (BOLD) signals], can be
measured and recorded with spatial resolution on the
order of 2--4 millimeters, taken every 2--4 seconds.
Noise reduction, image registration, outliers, image
detection, spatial and time trends, and multiplicity
are only some of the problems that can arise with
these data.  Among the first statisticians to attack
these problems were Keith Worsley and Karl Friston
[Worsley and Friston (\citeyear{WorFri95}); Worsley et al. (\citeyear{Woretal96}); Friston et al. (\citeyear{FriHol})] and
William Eddy and his colleagues [Eddy et al. (\citeyear{Eddetal95}); Eddy, Fitzgerald and Noll (\citeyear{EddFitNol96})],
who had sufficient computational resources at the time
to handle the massive amounts of data.  Since then,
computational power has significantly advanced,
enabling statisticians to investigate other
aspects of these types of data.  In addition, other
imaging methods have been developed with increased
sensitivity and resolution.  The first three articles in this section develop methods
   for analyzing fMRI data: Shi et al. (1), Vu et al. (2),
   and Bhattacharya and Maitra (3).
   %The methods in the next
%   three articles are developed in response to the availability
%   of data
Three articles develop methods for analyzing data
using more sensitive imaging techniques:
   Long et al. (4) model electromagetic source imaging data
   (magnetoencepholography imaging, or MEG); %Olhede and
%   Whitcher (7) analyze brain images from measurements obtained
%   via a particular type of magnetic resonance imaging known as
%   high angular resolution diffusion imaging (HARDI); and
%   Mishchencko et~al. (5) develop neural connectivity models
%   from data using calcium fluorescent imaging.
Mishchencko et al. (5) develop neural connectivity models from data using
calcium fluorescent imaging; and Olhede and Whitcher (7) analyze brain images
from measurements obtained via a type of magnetic resonance imaging known as
high angular resolution diffusion imaging (HARDI).
Neural spike trains collected from multielectrode
recordings motivate the methods in Kass et~al.~(6).

Shi et al.~(1) develop an \textit{adjusted exponentially
tilted empirical likelihood} method to detect differences
in the morphological changes, measured via fMRI, in
specific regions of the brain between two groups of
patients on different treatment protocols.  Beyond the
development of an appropriate model that accounts for
longitudinal measurements with time-varying covariates
is the challenge of developing a computational
algorithm to handle the data on 238 patients.
The results indicate regions of important differences
which provide insights into the different
mechanisms of the two treatment protocols.
Vu et al.~(2) use exploratory data analysis and
model selection procedures to improve a previously
proposed model for brain activity in encoding and
decoding sensory stimuli in the form of local constant
energy features.  Their analysis reveals nonlinearities
which, when incorporated into the model, yields a 25\%
improvement in encoding prediction and hence greater
accuracy in image identification.
Bhattacharya and Maitra (3) also analyze fMRI signals to
model dynamic, nonstationary neural connectivity via
a first-order vector autoregressive model which,
when applied to fMRI data on patients performing
specific tasks, provides insights into those brain
mechanisms involved in distinguishing shapes.

Data from more sensitive and higher resolution imaging
      techniques require more computationally intensive approaches.
Long et al. (4) develop
high-dimen\-sional (in the number of parameters)
state-space models to identifying magnitudes and
locations of neural sources that give rise to MEG signals
recorded on the surface of the head.  Due to the greatly
increased resolution of the data and the number of
parameters to be estimated, the Kalman filter solution
can be implemented only on high-performance
supercomputers.  The authors' Kalman filter
approach can be viewed as a specific implementation
of a more general approach using random field theory
proposed by Taylor and Worsley (\citeyear{TayWor07}) and applied to
MEG (and electroencepholography, or EEG) data by
Kilner and Friston (\citeyear{KilFri10}) that appeared in
\textit{The Annals of Applied Statistics} last year.

 The next two articles in this special section use
  different sources of data to model neuronal connectivity.
  One source of data is calcium-sensitive fluorescent
  imaging, which offers much finer spatial and temporal
  resolution than is possible with fMRI.
  Mishchencko et~al. (5) use such imaging data to model neural
  circuitry with a collection of coupled Hidden Markov models
  (HMMs), where each Markov chain represents the behavior of a
  single neuron and the coupling between the HMMs reflects
  the network connectivity matrix.  As is the case with the
  other articles in this section, the vast amounts of data
  and the complexity of the coupled models require clever
  computational approaches (in this case, a blockwise
  Gibbs algorithm) to estimate model parameters with
  biologically meaningful relevance.  %In the last of these
%  seven articles,
Kass et al. (6) consider models for data
  from external electrodes on the brain.  In the past,
  neural spike trains from external electrodes
  have been analyzed traditionally as point processes
  [Brillinger (\citeyear{Bri88}, \citeyear{Bri92})].  Such models usually assume
  stationarity and distinct events (no two events occur
  at the same time).  %Kass et al. (6)
  Here, Kass, Kelly, and Loh enhanced these
  models for neural spike trains by introducing a class
  of continuous-time-varying loglinear models which
  incorporates time-varying intensities,
  autocovariation, and synchrony.
  For an approach to estimating the number of neurons
  involved in a multi-neuronal spike train, see
  Li and Loh (\citeyear{MenLoh11}) that appeared in the most
  recent issue of \textit{AOAS}.

Olhede and Whitcher (7) approach the analysis of brain
  images through the local estimation of the two-dimensional
  probability density function (pdf) of HARDI measurements
  (i.e., measurements of the local molecular diffusion of
  water molecules, obtained via high angular resolution
  diffusion imaging).  Rather than assuming a Gaussian pdf,
  Olhede and Whitcher use the increased sampling rate of
  HARDI to estimate a nonparametric pdf using local
  measurements of the covariance matrix, enabling greater
  accuracy (less bias) at relatively little cost in terms
  of precision (increased variance).  However, because
  the data come from a diffusion process, the measurements
  are inherently spectral in nature.  The authors provide
  the statistical framework for estimating pdfs in the
  spectral domain, incorporating known properties of the
  diffusion process, and then use properties of Fourier
  transforms to invert the estimated pdf into the brain
  image domain.  Nonparametric tests for non-uniformity,
  asymmetry, and ellipsoidality in the pdf lead to increased
  understanding of diffusion in the brain.

 % Neural spike trains from external electrodes
%have been analyzed traditionally as point processes
%[Brillinger (\citeyear{Bri88}, \citeyear{Bri92})].  Such models usually assume
%stationarity and distinct events (no two events occur
%at the same time).  Kass et al.~(6)
%enhanced these traditional models for neural spike trains
%by introducing a class of continuous-time-varying loglinear
%models which incorporates time-varying intensities,
%autocovariation, and synchrony.
%Calcium-sensitive fluorescent
%indicators enable neural activity at much finer spatial
%and temporal resolution, thereby enabling the development
%of models for neuronal connectivity.
%Mishchencko et al.~(5) use such data to model neural
%circuitry with a collection of coupled Hidden Markov models
%(HMMs) where each Markov chain represents the behavior of a
%single neuron and the coupling between the HMMs reflects
%the network connectivity matrix.  As is the case with the
%other articles in this section, the vast amounts of data
%and the complexity of the coupled models require clever
%computational approaches (in this case, a blockwise
%Gibbs algorithm) to estimate model parameters with
%biologically meaningful relevance.  For an approach to
%estimating the number of neurons involved in a multi-neuronal
%spike train, see Li and Loh (\citeyear{MenLoh11}) that appeared in the most
%recent issue of \textit{AOAS}.

As Peter Hall indicated with respect to data in other fields,
here the analysis of neuroscience data led to
the development of new statistical methodology.
Besides the common theme of neuroscience as the
motivation for the methodology, all nine articles (the
present seven in this issue and the two articles that
appeared earlier) share two additional features:
(1) the analysis of very large data sets, which thereby
require (2) the development of computational algorithms
to facilitate estimation of complex models needed to
incorporate the nonstandard features of the data
(e.g., nonlinearity, nonstationarity, etc.).
Many more problems posed by these sorts of data are in
need of solutions, for example, relaxing assumptions on models,
designing experimental strategies to make best use
of the data, developing methods to reduce noise
(increase signal-to-noise ratio), etc.  Useful, practical
solutions can be obtained only through collaboration
between scientists and statisticians.
We hope that these articles will stimulate statisticians
and neuroscientists to collaborate on these problems to
further research in both domains.

% imsref loaded by dianan, 2011-05-09 16:06:35

\printaddresses

\end{document}